\documentstyle[revtex]{aps}

\newcommand{\be}{\begin{equation}}
\newcommand{\ee}{\end{equation}}
\newcommand{\bdm}{\begin{displaymath}}
\newcommand{\edm}{\end{displaymath}}
\newcommand{\ba}{\begin{eqnarray}}
\newcommand{\ea}{\end{eqnarray}}

\begin{document}

\begin{title}
Thin Films of $^3He$ --  Implications on the Identification of
$^3 He -A$
\end{title}

\author{S. K. Yip}

\begin{instit}
Department of Physics and Astronomy,
Northwestern University,
2145 Sheridan Road,
Evanston, IL 60208, U. S. A.
\end{instit}


\begin{abstract}

Recently the identification of $^3He-A$ with the axial state
has been questioned. It is suggested that the A-phase can actually be in the
axiplanar state.   We point
out in the present paper that experiments in a film geometry may be useful
to distinguish the above two possibilities. In particular a second order
phase transition between an axial and an axiplanar state would occur as
a function of thickness or temperature.


\end{abstract}

\vskip 1 cm


It has been common in the $^3 He$ literature that the `A-phase'
is identified as the `axial' or the `Anderson-Brinkman-Morel' state,
where, for suitably chosen coordinate axes,
 the orbital part of the pair wavefunction has the form of
$Y_1^1 (\hat k)$ of the spherical harmonics.  Recently, however, this
identification has been questioned \cite{tang}.  In particular,
it is pointed out that there is no strong evidence that the
`$\beta$-parameter' $\beta_{45}$ (for more details, see below)
is negative, and hence in principle the A-phase can actually
be in the `axiplanar' state.  The easiest way to visualize
this state is the following:  for a suitable choice of spin quantization
axes (not necessarily related to the axes for the orbital
wavefunctions below),
 there exist only $S_z = \pm 1 $ pairs; both kinds of pairs have their orbital
wavefunctions of the form $Y_1^1 (\hat k)$, yet with their own orbital
quantization axes, i.e., the two sets of orbital axes differ from each
other and thus there is a finite
angle between the `angular momentum' directions for the up and down pairs.
When these angular momentum axes coincide, the axiplanar state reduces to
the axial state.

In this note we point out that the effect of a smooth surface is very
different depending on whether the superfluid is in the axial or
the axiplanar state.  The intuitive picture is simple.  A smooth
surface allows the existence of the components of the orbital
order parameter parallel to the surface, whereas the other components
are suppressed \cite{a-dg-r}.
  Thus for a sufficiently confined geometry (thin film, or correspondingly,
sufficiently high temperatures)
the order parameter has to reduce to the axial state
(the planar state is unstable with respect to the axial state).
Experiments in a film geometry can thus be used to put constraints
on $\beta_{45}$.  For pressures
above the polycritical pressure,
if $\beta_{45}$ is indeed positive,  then while
in the bulk the axiplanar state is more stable than the axial state,
in a film as the temperature is lowered the
$^3He$ should make a phase transition from the normal state first
into the axial state then finally the axiplanar state (the second
transition is possible only if the film
is thick enough).
 For pressures below the polycritical pressure, as
the temperature is decreased
the $^3He$  can therefore evolve from the normal phase to the axial state
and then either directly,
or via the axiplanar state, to the B-planar phase depending on the value of
 $\beta_{45}$.

We illustrate this idea by finding the critical thickness
(or temperature) for the transition between the
axial and axiplanar states in a film with smooth
surfaces in Ginzburg-Landau theory.
We write the free enery as (cf. \cite{ho})
\bdm
f = f_c + f_g
\edm
where $f_c$ is the condensation energy density
\ba
f_c & =& - a_{\mu i} a^{\ast}_{\mu i} +
      \zeta_1 a_{\mu j} a_{\mu j} a^{\ast}_{\nu i} a^{\ast}_{\nu i}
    +  \zeta_2 a_{\mu i}  a^{\ast}_{\mu i} a_{\nu j} a^{\ast}_{\nu j}
   +  \zeta_3 a_{\mu j} a_{\nu j} a^{\ast}_{\nu i} a^{\ast}_{\mu i}
                       \nonumber \\
  & &   \qquad
 +  \zeta_4 a_{\mu j} a^{\ast}_{\nu j} a_{\nu i} a^{\ast}_{\mu i}
  +  \zeta_5 a_{\mu j} a^{\ast}_{\nu j} a^{\ast}_{\nu i} a_{\mu i}
\ea
and $f_g$ is the gradient energy density
\be
  f_g = {3 \over 5}
     \bigl( 2 (\partial_j a_{\mu j}) (\partial_i a^{\ast}_{\mu i})
   +  (\partial_i a_{\mu j}) (\partial_i a^{\ast}_{\mu j}) \bigr)
\ee
where we have normalized the order parameter
$a_{\mu i}$ and the free energy
 to those of the axial state
(i.e.  for the axial state $ a_{\mu i} = \hat d_{\mu}
 ( \hat m+ i \hat n)_i $ where $\hat d, \hat m, \hat n$ are all unit
vectors, and $ f_{c} = -1 $ ) and have simply adopted the
weak-coupling form of the gradient energy.
The lengths are in units of the temperature dependent coherence length
$\xi (T) = \xi_{o} / (1 - {T \over T_c^o})^{1 / 2}$,
$ \xi_{o}^2 = { 7 \zeta (3) \over 48 \pi^2 }
         ( { \hbar v_f \over k_B T_c^o})^2  $.
Here $T_c^{o}$ is the bulk transition temperature.
The $\zeta$ parameters are related to the usual $\beta$ parameters
via $ \zeta_i = \beta_i / (4 \beta_{245})$.  We consider a film
in the x-z plane of real thickness $D$ (dimensionless
thickness $ d \equiv D / \xi(T)$ ).   In the thin film limit
$a_{\mu i} \not= 0 $ only for $ i=x$ or $z$.
We then only need to consider order parameters of the form
\be
a_{\mu i} =
\left (
\begin{array}{ccc}
0    & a_{xy}   &  0 \\
a_{yx}  & 0     & a_{yz}   \\
0 &     0       &  0
\end{array}
\right )
\ee
For the axiplanar state in the bulk $a_{\mu i}$ then has the form
given by Barton and Moore \cite{barton}.  In the thin film
(high temperature) limit we can choose $ a_{yx} = 1, a_{yz} = i$.
 (We can imagine
a magnetic field in the $\hat z$ direction so that $\hat d = \hat y$.)
Variation of the free energy with respect to $a^{\ast}_{xy}$ leads to
the differential equation
\ba
0 &=& - { 9 \over 5} \partial_y^2 a_{xy} - a_{xy}
    + 2  \zeta_1 a^{\ast}_{xy} ( a_{yx}^2 + a_{yz}^2 + a_{xy}^2 )
          \nonumber  \\
  & & \qquad  +  2  \zeta_2 a_{xy} (\vert a_{yx} \vert^2 +
            \vert a_{yz} \vert^2 + \vert a_{xy} \vert^2 )
   + 2 \zeta_{345} a^{\ast}_{xy} a_{xy}^2
\ea
Thus as expected the transition between the axial and axiplanar states,
if exists, is second order.  The highest transition temperature
is for $a_{xy}$ of the form $a^{o} {\rm sin} { \pi y \over d}$
so that $a_{xy}$ vanishes at both $y=0$ and $y=d$ as required.
The transition temperature is obtained by setting the coefficient of
$a^{o}$ to zero, i.e.,
\be
 -1 + { 9 \pi^2 \over 5 d^2 } + 4 \zeta_2  = 0
\ee
Thus the transition exists only when $\zeta_2 < { 1 \over 4}$,
i.e. $ \beta_{45} > 0 $, as it must be.  In ordinary units
the transition temperature $T_c$ is given by
\be
d_{axi}^2 \equiv D^2 / [ \xi_o^2 ( 1 - {T_c \over T_c^{o}})]
   = { 9 \pi^2 \over 5} {\beta_{245} \over \beta_{45} }
\ee

We would like to deduce the restrictions on $\beta_{45}$ from
the existing experiments in the film geometries so far.
At least two complications have to be kept in mind:
(i) the surfaces involved
thus far are probably not smooth; and (ii) the above calculation is only
valid within the Ginzburg-Landau regime.  Modifying the above
calculation for rough surfaces can in principle be done, though
since the order parameter even in the axial state would be
non-uniform numerical calculations need to be performed.
Searching for the phase transition beyond the Ginzburg-Landau
region would require a microscopic theory which includes
properly the strong coupling effects.  These calculations
would not be performed here.  Thus the comparison with experiments
below should be regarded as for illustrative purposes only.

At pressures below the polycritical and sufficiently large
$d$ the order parameter should be in the $B$-planar state. \cite{ho}
Experiments so far have yielded only either
no transition \cite{freeman1,freeman2}
or one (first order) transition \cite{harrison89,crooker},
but not two. Assuming that this implies that the axiplanar phase
has not existed in films so far investigated yields
\bdm
d_{axi} > d_{expr} \qquad {\rm or} \qquad d_{axi} > d_{B}
\edm
where $d_{expr}$ are the values of $D/\xi(T)$ covered in
experiments and $d_B$ is the critical $D/\xi(T)$ for the transition
between the axial state and the B-planar state.
[Actually the experiment by Xu and Crooker \cite{crooker}
 covered a region of $d$ somewhat smaller
than the value of $d$ at which Harrison et. al. \cite{harrison89} observed
their phase transition in the corresponding experiment (i.e., with no
preplated $^4He$.) Moreover the phase transition in \cite{crooker}
seems to occur at a fixed {\it real} thickness $D$.
It is unclear whether both these (first order )
transitions corresponds to the
B-planar $ \rightarrow $ axial states
transition, or they are actually different.]
These imply upper limits for $\beta_{45}$:
\bdm
\beta_{45} < { 9 \pi^2 \over 5} {\beta_{245} \over d_{expr}^2 }
\edm
and
\bdm
\beta_{45} < { 9 \pi^2 \over 5} {\beta_{245} \over d_{B}^2 }
\edm
respectively.
The zero pressure experiments \cite{harrison89,crooker}
indicating transitions at relatively small $d$ ( $\sim 3-5$ ),
and the higher pressure experiments of Freeman et. al.
\cite{freeman2} at
relatively high temperatures,
do not impose a stringent condition on $\beta_{45}$.
The experiment at 9 bar by Freeman et. al. \cite{freeman1,freeman2},
for thickness of $D \approx 3000 \AA $ with minimum
$T / T_c^o \approx 0.2 $, fails to see the B-planar state.
This implies $d_{B} { > \atop \sim} 12$ and thus $\beta_{45} / \beta_2
{< \atop \sim} 0.13 $.  Clearly more restrictive conditions
(and more relevant for the sign of $\beta_{45}$ above
the polycritical pressure) can be given if experiments are
carried out at higher pressures (where $d_B$ is expected to
increase) and at lower temperatures.

Experiments for pressures above the polycritical are more limited,
and to my knowledge no transition has been seen.
This may imply for all experiments so far $d_{expr} < d_{axi}$,
and all the films investigated so far are in the axial
state.  The $22$ bar experiment of Freeman et. al \cite{freeman2}
implies roughly $ \beta_{45}/ \beta_2 {< \atop \sim} 0.17 $.
The above restrictions that can be placed on $\beta_{45}$,
however, seemingly are not as strong as those based on
the existing NMR experiments
\cite{rand}.

In conclusion we have pointed out that there is a crucial difference
in the behavior of the axial and axiplanar states near a
smooth surface, and experiments in film geometries can be useful
in distinguishing which state is actually realized in
superfluid $^3 He-A$.


\bigskip

This work was supported by the National Science Foundation through the
Northwestern University Materials Science Center, Grant No. DMR
8821571, DMR-9120521.


\begin{thebibliography}{9}

\bibitem{tang} Y. H. Tang, I. Hahn, H. M. Bozler and C. M. Gould,
  {\it Phys. Rev. Lett.} {\bf 67}, 1775 (1991).

\bibitem{a-dg-r} V. Ambegaokar, P. G. de Gennes and D. Rainer,
  {\it Phys. Rev. } {\bf A9} 2676 (1974).

\bibitem{ho}
	Y-H. Li and T-L. Ho, {\it Phys. Rev. } {\bf B38 } 2362 (1988).

\bibitem{barton}
	G. Barton and M. A. Moore,
	{\it J. Phys. C.: Solid State Phys.},
	{\bf 7}, 4220 (1974).

\bibitem{freeman1}
	M. R. Freeman {\it et. el.},
 {\it Phys. Rev. Lett.} {\bf 60}, 596 (1988).

\bibitem{freeman2}
	M. R. Freeman and R. C. Richardson,
     {\it Phys. Rev. B } {\bf 41} 11011 (1990).

\bibitem{harrison89}
	J. P. Harrison {\it et el.},
	{\it Quantum Fluids and Solids -- 1989},
      ed. by G. G. Ihas and Y. Takano,
	American Institute of Physics Conf. Proceedings 194, 221 (1989)

\bibitem{crooker}
	J. Xu and B. C. Crooker,
  {\it Phys. Rev. Lett.} {\bf 65}, 3005 (1990).

\bibitem{rand}
	M. Rand {\it et. al.},
        {\it Proc. of 20th International Conference on Low Temperature
              Physics, 1993}





\end{thebibliography}
\end{document}